\documentclass[10pt,conference]{IEEEtran}
\usepackage{cite}
\usepackage{amsmath,amssymb,amsfonts}
\usepackage{algorithmic}
\usepackage{graphicx}
\usepackage{textcomp}
\usepackage{xcolor}
\usepackage[hyphens]{url}
\usepackage{fancyhdr}
\usepackage{hyperref}
\usepackage{algorithm}
\usepackage{algorithmic}
\usepackage{comment}

\title{FIITED: Fine-Grained Embedding Dimension Optimization During Training for Recommender Systems}
\makeatletter
    \newcommand{\linebreakand}{%
      \end{@IEEEauthorhalign}
      \hfill\mbox{}\par
      \mbox{}\hfill\begin{@IEEEauthorhalign}
    }
\makeatother

\IEEEoverridecommandlockouts

\author{\IEEEauthorblockN{\textbf{  Qinyi Luo*\textsuperscript{$\dagger$} \thanks{*Equal contribution.} 
     \thanks{\textsuperscript{$\dagger$}Qinyi Luo completed part of this paper during her internship at Meta.}}}
    \IEEEauthorblockA{
    University of Southern California \\
    Los Angeles, CA, USA}
    \and
    \IEEEauthorblockN{\textbf{    Penghan Wang*\textsuperscript{$\ddagger$} \thanks{\textsuperscript{$\ddagger$}Penghan Wang participated in this paper during his internship at Purdue University.}}}
    \IEEEauthorblockA{
    Tsinghua University \\
    Beijing, China}
    \and
    \IEEEauthorblockN{\textbf{Wei Zhang}}
    \IEEEauthorblockA{
    Meta \\
    USA}
    \and
    \IEEEauthorblockN{\textbf{  Fan Lai\textsuperscript{$\mathsection$} \thanks{\textsuperscript{$\mathsection$}Fan Lai participated in this paper as a visiting researcher at Meta.}}}
    \IEEEauthorblockA{
    University of Illinois, Urbana-Champaign \\
    Urbana, IL, USA \\}
    \linebreakand % <------------- \and with a line-break
    \and
    \IEEEauthorblockN{\textbf{Jiachen Mao}}
    \IEEEauthorblockA{
    Meta \\
    USA}
    \and
    \IEEEauthorblockN{\textbf{Xiaohan Wei}}
    \IEEEauthorblockA{
    Meta \\
    USA}
    \and
    \IEEEauthorblockN{\textbf{    Jun Song\textsuperscript{$\mathparagraph$} \thanks{\textsuperscript{$\mathparagraph$}Jun Song participated in this paper during her internship at Meta.}}}
    \IEEEauthorblockA{
    University of Washington \\
    Seattle, WA, USA}
    \and
    \IEEEauthorblockN{\textbf{Wei-Yu Tsai}}
    \IEEEauthorblockA{
    Meta \\
    USA}
    \and
    \IEEEauthorblockN{\textbf{Shuai Yang}}
    \IEEEauthorblockA{
    Meta \\
    USA}
    \and
    \IEEEauthorblockN{\textbf{Yuxi Hu}}
    \IEEEauthorblockA{
    Meta \\
    USA}
    \and
    \IEEEauthorblockN{\textbf{Xuehai Qian}}
    \IEEEauthorblockA{
    Tsinghua University \\
    Beijing, China}
    }

\begin{document}
\maketitle

\thispagestyle{plain}
\pagestyle{plain}

%%%%%%%%%%%%%%%%%%%%%%%%%%%%%%%%%%%%%%%%
%%%%%%%% -- PAPER CONTENT STARTS -- %%%%%%%%%

\begin{abstract}

  Huge embedding tables in modern deep learning recommender models (DLRM) require prohibitively large memory during training and inference. This paper proposes FIITED, a system to automatically reduce the memory footprint via FIne-grained In-Training Embedding Dimension pruning. By leveraging the key insight that embedding vectors are not equally important, FIITED adaptively adjusts the dimension of each individual embedding vector during model training, assigning larger dimensions to more important embeddings while adapting to dynamic changes in data. 
We prioritize embedding dimensions with higher frequencies and gradients as more important. To enable efficient pruning of embeddings and their dimensions during model training, we propose an embedding storage system based on virtually-hashed physically-indexed hash tables. Experiments on two industry models and months of realistic datasets show that FIITED can reduce DLRM embedding size by more than 65\% while preserving model quality, outperforming state-of-the-art in-training embedding pruning methods. On public datasets, FIITED can reduce the size of embedding tables by 2.1$\times$ to 800$\times$ with negligible accuracy drop, while improving model throughput.

\end{abstract}

\section{Introduction}

Unlike computer vision (CV) and natural language processing (NLP) models, modern deep learning recommendation models (DLRMs) take both sparse categorical features (e.g., user IDs) and dense features (e.g., commodity price) as inputs to produce model output.
In a typical deep learning recommendation model, each sparse feature is represented by an entry row in the embedding table, and each row contains the embedding vector of the sparse feature. The size of an embedding table is determined by the number of rows (i.e., hash size), the number of columns (i.e., embedding dimension size), and the size of each value in the embedding vector.

As accuracy demands increase--where improvements greater than 0.1\% are deemed significant~\cite{zhou2018deep, song2019autoint}--modern DLRMs incorporate more sparse features and instances. This requires terabytes of memory across multiple GPUs to maintain the desired throughput~\cite{persia}. Unfortunately, current model size growth outpaces hardware advancements~\cite{persia}, and the ever-growing embedding size can lead to the underutilization of GPU compute resources. This is because embedding operations are memory-intensive rather than computation-intensive. As such, reducing the memory cost of embedding tables is crucial for efficient DLRM execution and allows for sustainable model development.

However, determining the optimal allocation of embeddings is challenging. Sparse features, thus their associated embedding tables, vary in importance to model output, with even rows within the same table having different importance due to heterogeneous characteristics (e.g., access patterns) and dynamics over time. 
Moreover, using exceedingly large dimensions of embedding rows not only amplifies their memory demands but can lead to model overfitting (e.g., due to insufficient training and data), while too small dimensions are 
insufficient to express information contained in the sparse features.
Despite recent works ~\cite{sseds,learn_elastic,mix_dim,esapn,pep,twin_layer,autodim,autoemb,field-wise,optemed,autosrh,i-razor}, setting embedding dimensions in real-world applications still heavily relies on empirical evidence, leading to great engineering effort and hard to adapt over time and across tasks.

\begin{figure*}
\begin{center}
\includegraphics[width=0.95\linewidth]{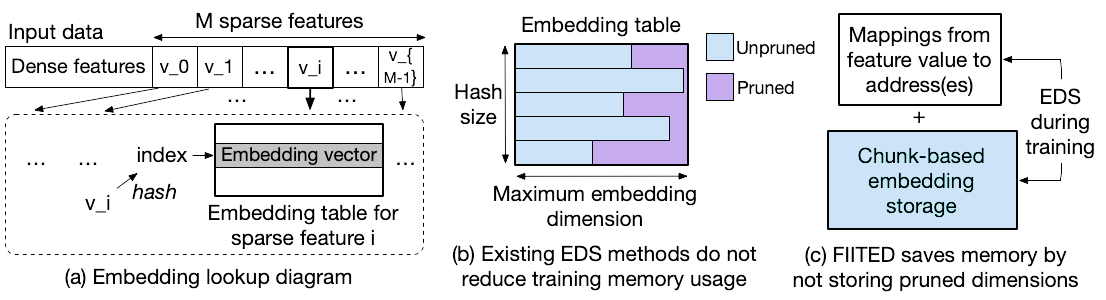}
\end{center}
\caption{Existing Embedding Dimension Search (EDS) methods vs. FIITED. 
}
\label{fig:intro}
\end{figure*}

In this paper, we propose FIITED, a system for FIne-grained In-Training Embedding Dimension optimization. FIITED allocates more memory to more important embeddings and prunes the dimensions of less important embeddings to make better use of limited memory resources, adjusting the dimension of each embedding vector during training. Important embeddings are identified via importance scores computed during training which are based on embedding characteristics including access frequency and gradient norms. As a result, FIITED can reduce both the model size and the memory footprint
during training to a desired amount.

Compared with existing embedding dimension search (EDS) efforts ~\cite{sseds,mix_dim,learn_elastic,pep,autodim,field-wise,optemed,autosrh,i-razor}, 
the main advantage of FIITED, as shown in 
Figure~\ref{fig:intro}, is enabling automated memory saving during training by not storing embeddings at the maximum length.
Reducing the memory consumption of training is highly beneficial, as it enables training huge DLRMs on hardware devices with low on-device memory (e.g., GPUs)~\cite{merlin}, lessens the need to spend extra time transferring embeddings from off-device memory-rich hardware (e.g., CPUs and SSDs)~\cite{recshard,SSD}, and potentially frees up memory for more features to be added to the model, which can improve the model quality. Moreover, by automatically identifying the right embedding dimension allocation, FIITED can reduce laboring efforts while achieving even better model accuracy.
However, it is non-trivial to realize, because naively pruning embedding dimensions will result in many tiny fragments of free memory which are hard to utilize.
To tackle this challenge, we propose a novel chunk-based embedding storage system.

In addition to reducing the training memory footprint, FIITED brings three more
advantages compared to most of the existing EDS methods. First, FIITED is fined-grained and adjusts the embedding dimension at the embedding vector level, while most EDS methods operate at the sparse feature level and set a uniform dimension for
all embeddings within the same sparse feature ~\cite{sseds,learn_elastic,mix_dim}. Fine-grained EDS is necessary because even within the same sparse feature, some embeddings may be more important than others and thus require different dimensions~\cite{adaembed,esapn}.
Second, FIITED performs EDS \textit{during training} and can take advantage of dynamic changes in data characteristics. Since the importance of an embedding vector often changes over time~\cite{adaembed,esapn}, it is natural to assume that their optimal dimension should vary accordingly. In-training EDS offers unique opportunities to adjust embedding dimensions over time during training, which is not allowed in pre-training or post-training EDS methods ~\cite{sseds,mix_dim,learn_elastic, autodim, autosrh}. Moreover, FIITED does not rely on any prior knowledge of the training data statistics but rather adapts to data characteristics during training, and therefore is better suited to
application domains with fast-changing data traits or when training data are not fully observable at the start of training, e.g., during online learning.
Third, FIITED does not need any pre-training or re-training procedure, and thus has relatively low training time overhead.

In summary, this paper makes three main contributions:

\begin{itemize}
    \item We propose FIITED, a novel fine-grained embedding dimension optimization method that adjusts the dimension of each embedding vector during training. FIITED directly cuts down training memory footprint by adopting a chunk-based embedding storage system design.
    \item The proposed method performs fine-grained EDS, adapts to changing data characteristics over time, and does not require pre-training, re-training, or prior knowledge of training data traits.
    Thus, FIITED is more flexible, faster, and easier to use than most of the existing EDS methods, while achieving a higher reduction in the model size without hurting model quality.
    \item Experiments on two industry models show that FIITED can reduce a significant amount of embedding size during training ($>$65\% and up to 50\% for the two models, respectively) while maintaining the quality of the trained model. Compared to a state-of-the-art in-training embedding pruning method, AdaEmbed~\cite{adaembed}, FIITED is able to achieve higher pruning ratios without affecting model quality. On public datasets, FIITED is able to reduce the embedding table by 100$\times$ to 800$\times$ on three click-through-rate datasets and $2.2\times$ to $17.8\times$ on one classification dataset, with little accuracy drop.
\end{itemize}

\begin{figure*}
%\vskip 0.2in
\begin{center}
\centerline{\includegraphics[width=\linewidth]{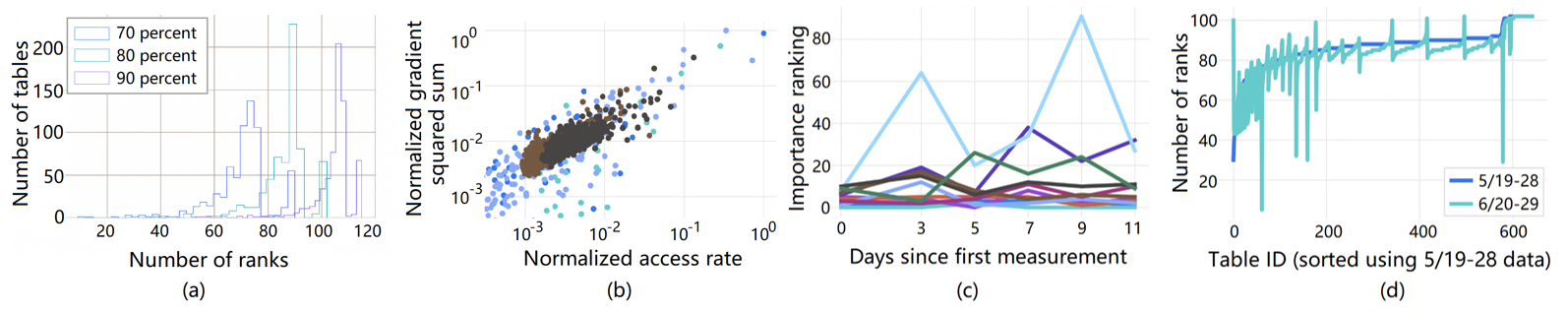}}
\caption{Embedding characteristics. (a) Histogram of the number of ranks needed to preserve a certain percentage of squared
SVD values for embedding tables. (b) Normalized row access and gradient squared sum of randomly sampled rows from 10
biggest tables. Both axes are log-scale. (c) Importance rankings of the top 10 important features change over time. (d) SVD characteristics change over time. The average absolute change is 4.4.
}
\label{fig:back1}
\end{center}
\vskip -0.2in
\end{figure*}

\section{Background and Motivations}\label{sec:back}

In this section, we first briefly give an background introduction to DLRMs, and then analyze the characteristics of sparse feature embedding table observed during our experiment, which leads to the motivation of our solution.

\subsection{Deep Learning Recommendation Models}\label{sec:back_DLRM}

The input of a DLRM can be categorized into sparse features (e.g. post IDs, user IDs) and dense features (e.g. timestamps). A deep learning recommendation model (DLRM) mainly consists of two parts, an embedding table for sparse features and multiple-layer perceptron (MLP) neural networks for dense features. For dense features, the features are fed into a MLP neural network to capture the weight vectors of the features. For sparse features, the embedding vectors need to be retrieved from the embedding table first, where each row contains the embedding weight vector of a sparse feature. The retrieved embedding vectors of sparse features are then interacted (e.g. concatenated or element-wise multiplied) with the weight vectors of dense features. The interaction results are then used for downstream tasks. For example, in click-through rate prediction tasks, the result vectors are fed into another MLP neural network to produce the final prediction result. Modern DLRMs normally have a large number of sparse features as a result of the enormous number of instances. Consequently, the embedding table causes the majority of memory consumption in DLRM systems, and a reduction in the size of the embedding table can lead to a reduction in the memory footprint of the entire model.

\begin{figure}
%\vskip 0.2in
\begin{center}
\centerline{\includegraphics[width=\columnwidth]{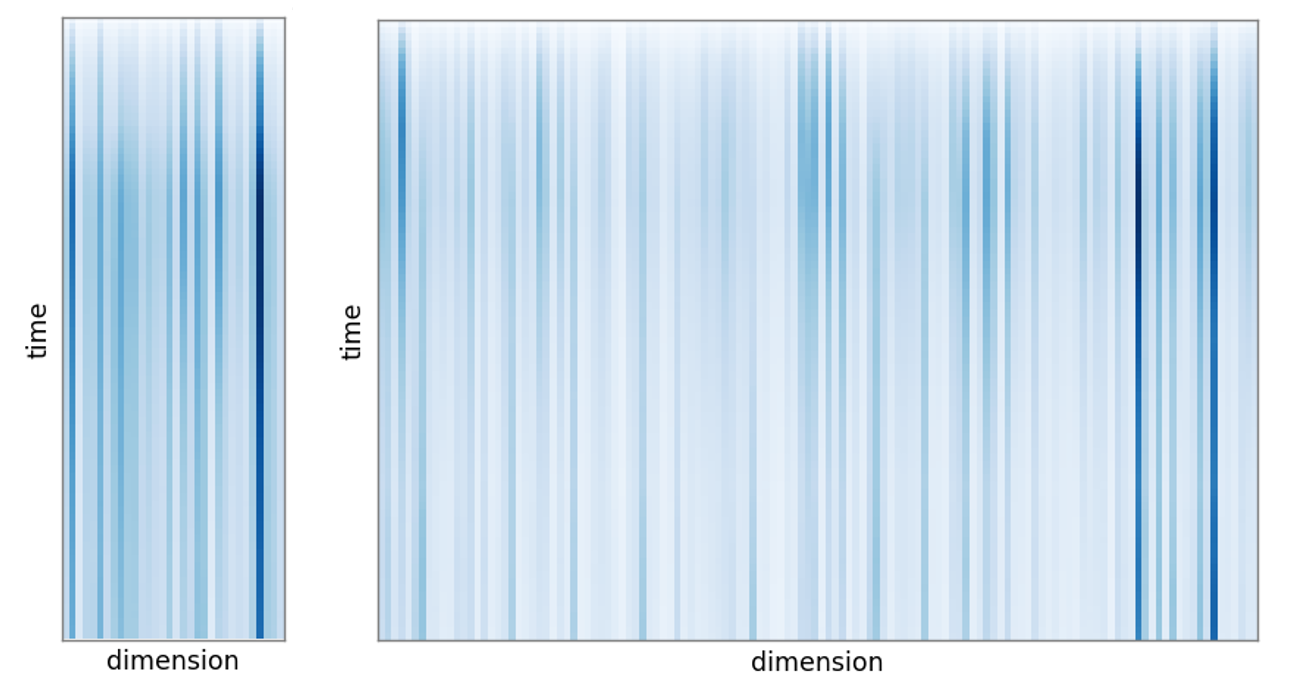}}
\caption{Change of dimension importance on two DLRM systems with different dimension sizes. The deeper color indicates a larger importance value.}
\label{fig:dimension_importance}
\end{center}
\vskip -0.2in
\end{figure}

\subsection{Embedding Table Characteristics}\label{sec:characteristics}

In this section, we analyze the characteristics of sparse features and their embeddings, which serve as the motivation for our in-training embedding dimension optimization method. A trained private DLRM model was used for the analysis.

\textbf{Heterogeneous Importance of Sparse Features.}
For each sparse feature's trained embedding table, we compute the number of ranks required to keep a certain percentage of squared SVD values, which serves as a proxy of the amount of information in the table, and plot a histogram across all tables. The histogram, as seen in Figure~\ref{fig:back1}(a), shows that the number of ranks varies substantially for different tables and covers a large range.

\textbf{Heterogeneous Importance of Embeddings.}
Rows within the same table are not equally important either. We examine the largest 10 embedding tables in the DLRM and sample 1k rows randomly from each of them, and plot the normalized row access frequency and gradient squared sum, as shown in Figure~\ref{fig:back1}(b). It can be seen that both metrics span a wide range, with a tiny fraction of rows showing significantly larger values. These ``hot'' rows are accessed much more frequently, resulting in a much higher gradient squared sum. Because they have been updated more, they have better quality and thus can likely contribute more to the model quality.

\textbf{Heterogeneous Importance across Dimensions.}
Upon previous insights, we further explored the importance of embedding tables across all dimensions, where some dimensions are inherently more important than other dimensions on all entries. The importance of a specific dimension is calculated using the gradient squared sums of all embedding entries on that dimension. Figure~\ref{fig:dimension_importance} shows the change of importance of each dimension during training time, sampled from two DLRM systems with varied embedding dimensions. It can be seen that some dimensions consistently remain important, while some dimensions bear lower importance scores during the entire training time.

\begin{figure}
%\vskip 0.2in
\begin{center}
\centerline{\includegraphics[width=\columnwidth]{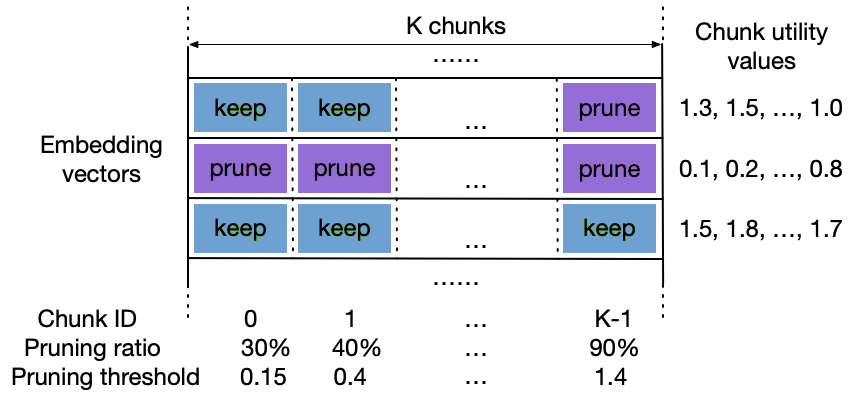}}
\caption{In-training embedding dimension pruning.}
\label{fig:example}
\end{center}
\vskip -0.2in
\end{figure}

\begin{figure*}
%\vskip 0.2in
\begin{center}
\centerline{\includegraphics[width=0.6\linewidth]{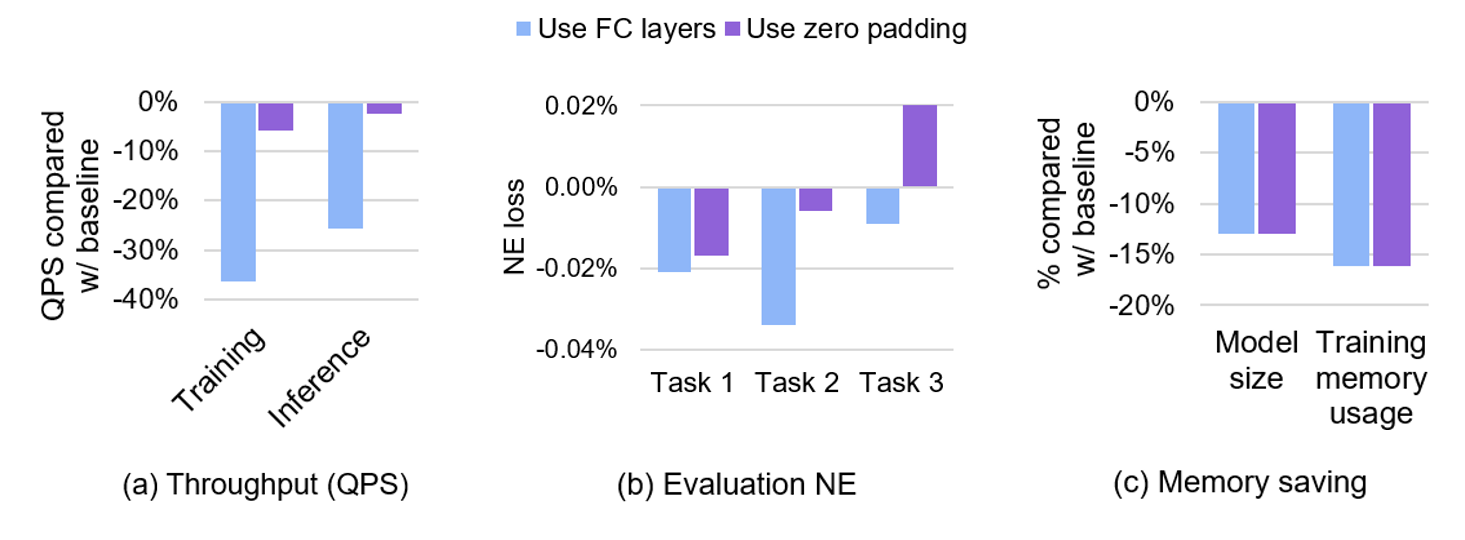}}
\caption{Zero padding vs. linear projection given the same per-feature embedding dimensions.}
\label{fig:zero}
\end{center}
\vskip -0.2in
\end{figure*}

\textbf{Change of Feature Characteristics Over Time.}
The characteristics of features and of their embeddings also change over time. 
In Figure~\ref{fig:back1}(c), we show how feature importance changed over a duration of 12 days. Here, the importance of a sparse feature is computed by first randomizing its embedding table and then computing how much model accuracy is decreased. 
We choose the 10 most important features on the beginning day and plot their importance rankings in the next 11 days. The figure shows that
some features’ importance rankings changed significantly from the top 10 to 20-40 or even above 80.

Another example of such changes can be seen in the SVD analysis of embedding tables. We compare SVD results of embedding tables trained over 2 different 10-day periods of data that are 1 month apart. For each table, we compute the number of ranks needed to retain 80\% of the squared SVD values, which is a proxy of how much information is contained in a table. We can see from Figure~\ref{fig:back1}(d) that the number of ranks is quite different between the two 10-day periods, meaning that the amount of information contained in each table can change over time as well.

\section{In-Training Embedding Dimension Optimization}

In order to set appropriate embedding dimensions, one needs to capture the amount of information contained in embedding vectors, which varies across sparse features, across embeddings of the same sparse feature, and changes over time~\cite{adaembed,esapn}
. To tackle this challenge, we design an in-training embedding dimension optimization method that dynamically adjusts the dimension of each embedding vector during training. 

\subsection{FIITED Overview}\label{method_overview}

Aiming to optimize embedding dimensions (i.e., dimension pruning), FIITED is built upon a state-of-the-art in-training embedding pruning method, AdaEmbed~\cite{adaembed}, which prunes entire embedding vectors during training (i.e., row pruning; each row in an embedding table stores one embedding vector). AdaEmbed assigns a row
utility value to each embedding row based on row access frequency and gradient information, and the rows are pruned periodically by comparing the utilities against a pruning threshold. The threshold is decided by a pre-defined pruning ratio $p$. The current utility values are sorted in ascending order and the value at the $p$-th percentile is selected as the threshold. The value $p$ can be selected by the user and indicates the desired total size of embeddings. Rows with utilities below the threshold are pruned, while the rest remain. 

To realize dimension optimization, we extend the row pruning method in AdaEmbed and divide every row in the
embedding tables into $K$ chunks, where each chunk is assigned its own pruning ratio. A utility metric is maintained during training for each chunk, and a total of $K$ pruning thresholds are computed instead of 1, with 1 threshold per chunk. Chunks with utilities below the corresponding threshold are
pruned. AdaEmbed becomes a special case of our dimension pruning method when $K=1$. As to how to decide the $K$ pruning ratios, we provide two methods: (1) manual selection, and (2) dynamic generation at runtime. Details of the two methods will be explained in Section~\ref{method_alg} and quantitative comparisons will be given in Section~\ref{exp}. 

To further explain the idea, an example is illustrated in Figure~\ref{fig:example}. Each row in the diagram
is an embedding vector with $K$ embedding chunks, and each chunk has its own utility value. Given pruning ratios
for the embedding chunks, a pruning threshold is computed for each chunk based on chunk
utility values. For each embedding chunk, the pruning decision is
made by comparing its utility against the pruning threshold. For example, in the first row,
the first chunk's utility 1.3 is bigger than the threshold 0.15, so it is retained; the last chunk has utility 1.0 which is smaller than the threshold 1.4, so it is pruned. A previously pruned chunk can be brought
back if its utility becomes larger than the threshold, and the chunk’s embedding values will be re-initialized.

\textbf{Zero padding VS. linear projection.} The fine-grained dimension pruning in FIITED results in different embeddings having different lengths, which poses a challenge for the computation 
during training and inference. In DLRMs, the dot
product is usually computed among embedding vectors to capture their interaction, and it
only works with embeddings of the same length. In previous works~\cite{sseds,mix_dim}, this problem is usually solved by inserting a fully connected (FC) projection layer for each table to map embeddings of shortened lengths to the maximum length, so that any two embeddings will have the same length during dot product.
But this strategy does not work for FIITED because even embeddings of the same sparse feature in the same table can have different lengths. Instead, we adopt zero padding to restore the embeddings
to their original length, i.e., pruned chunks are treated as all zeros. If a fully pruned row is fetched during training, an all-zero embedding is returned.

Since computing dot products on zero-padded embeddings can cause information loss due to multiplications by zero, it may seem that padding zeros will restrict the model quality. But it is not an issue: the idea
is that during training, the model can adjust itself and learn to accumulate information in the
unpruned chunks. To corroborate our claim, we performed preliminary tests that compare FC
layers with zero-padding during training of a Multi-Task-Multi-Label industry model. Prior to training, mixed embedding dimensions are assigned based on SVD analysis of a previously trained model, and they remain fixed during training. Model quality is measured by Normalized Entropy (NE)~\cite{NE1,NE2}, with lower values indicating better models. Results (plotted in Figure~\ref{fig:zero}) show that, compared to a baseline model with uniform embedding dimensions, zero padding only incurred a marginal NE loss
(0.020\%) in one of the three tasks while the other two tasks had tiny NE gains
(-0.017\%, -0.006\%), indicating that zero-padding is indeed a feasible approach.

\subsection{Dimension Optimization Algorithm}\label{method_alg}

The dimension optimization procedure is detailed in Algorithm~\ref{alg:fiited}. Chunk utility values are updated in every training iteration. The utility values can simply record the running average of row access frequency, or they can be designed to incorporate more complex information, e.g., the L2 norm of gradients. Our preliminary results show that incorporating gradient information can lead to better model accuracy.
Therefore, chunk utility in the $i$-th iteration is computed by 

\begin{equation}
    \label{eq:utility}
    u(i) = \gamma u(i-1) + a(i)g(i)
\end{equation}

where $a(i)$ is the number of accesses to the chunk in iteration $i$, $g(i)$ is the L2 norm of the gradients computed for the chunk in iteration $i$, and $\gamma \in (0,1)$ is a decay parameter that reduces the influence of history utility values in order to capture dynamic changes in data characteristics.
Using $a(i)g(i)$ as the utility metric has a theoretical grounding ~\cite{adaembed} in the training instance sampling problem, where such a metric is shown to accelerate convergence ~\cite{sampling_instance}, but instead of selecting training instances, we are selecting embedding chunks to train on.

Embedding pruning occurs every $T$ iterations. Empirically, $T$ is set to around one hour's worth of training data. During pruning, pruning thresholds are computed by first sorting the current utility values and then selecting the value at the desired pruning ratio. Instead of sorting all the utility values which can take a long time, a small number of utility values are randomly sampled. After obtaining the thresholds, FIITED will calculate the number of embedding chunks crossing the threshold, i.e. previously evicted chunks with utility values larger than the threshold or previously preserved chunks with utility values now smaller than the threshold. If this number exceeds a pre-defined ratio, an enforced pruning is executed, where embedding chunks with utility below the thresholds are pruned while the rest are retained in the model.

\begin{algorithm}
   \caption{In-training embedding dimension optimization}
   \label{alg:fiited}
\begin{algorithmic}
   \STATE {\bfseries Input:} pruning period $T$, sampling size $m$ 
   \FOR{{\bfseries training batch ID} $i$}
   \STATE Access embedding chunks, perform forward pass and backward pass
   \STATE Update embedding utility values
   \IF{$i$ \% $T$ = 0}
   \STATE Randomly sample $m$ utility values and sort them
   \STATE Compute pruning thresholds for embedding chunks
   \STATE Check if enforced pruning is needed for the pruning thresholds
   \IF{$enforce\_pruning == True$}
   \FOR{{\bfseries every embedding chunk} $e_j$}
   \IF{utility value $u_j$ is below threshold and $e_j$ is not pruned}
   \STATE Set chunk $e_j$ to pruned (i.e., evict $e_j$)
   \ENDIF
   \ENDFOR
   \ENDIF
   \FOR{{\bfseries every embedding chunk} $e_j$}
   \IF{utility value $u_j$ is above threshold and $e_j$ is pruned}
   \STATE Set chunk $e_j$ to not pruned and initialize chunk $e_j$ (i.e., allocate $e_j$)
   \ENDIF
   \ENDFOR
   \ENDIF
   \ENDFOR
   %\UNTIL{$noChange$ is $true$}
\end{algorithmic}
\end{algorithm}

\begin{figure*}
%\vskip 0.2in
\begin{center}
\centerline{\includegraphics[width=0.55\linewidth]{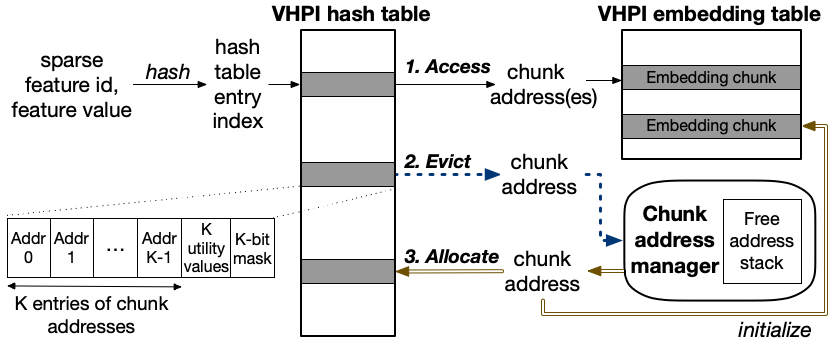}}
\caption{VHPI embedding table design that supports training memory saving in FIITED's chunk-based dimension pruning.}
\label{fig:system}
\end{center}
\vskip -0.2in
\end{figure*}

\textbf{Manual VS. adaptive per-chunk pruning ratios.} Individual pruning ratios for each chunk can be selected either manually or adaptively according to data characteristics. For manual selection, a pruning ratio $p_k$, $k=0,1,..., K-1$, for each chunk is selected empirically before training. For adaptive selection, a global pruning ratio $p$ is decided by the user and indicates the average pruning ratio across all the chunks. Utility values in all chunks are sorted together to decide one global pruning threshold. Individual pruning ratios for different chunks may be bigger or smaller than $p$, and may change over time, depending on chunk utility values computed at runtime. We evaluate both approaches in the experiments (Section~\ref{exp}).

\textbf{2D pruning.} Although our current dimension optimization algorithm already captures the characteristics of embedding tables in recommender systems, there is still further potential optimization due to the uneven property of dimension importance, as discussed in Section~\ref{sec:characteristics}. As a result, we further harness this characteristic by introducing a dimension-level pruning method on top of the current entry-level pruning method, forming a 2D pruning system together. In dimension-level pruning, we maintain the utility values of each dimension by calculating the squared gradient sum of all embedding entries on that dimension, and prune the dimensions using a similar approach as introduced above. Specifically, we only preserve the important dimensions of all entries in the embedding table, while the unimportant ones are replaced with zero. Through this mechanism, we can further push the boundary of pruning ratios on the basis of the current design. 

At the beginning of the training procedure, the entire embedding dimension is still needed to identify the hot dimensions, which we call a cold-start process. After the first dimension pruning decision is made, the embedding dimension is shrunk to the desired size, and only the preserved dimensions will be stored. Furthermore, as showcased in Section~\ref{sec:characteristics}, the hot dimensions, once identified, remain consistent during the whole training procedure. Hence, dimension pruning decisions only need to be made scarcely, or only during the cold-start period at the beginning of the training procedure, introducing little overhead to the current system. We offer dimension-level pruning as an optional addition to our system, where users can decide the settings based on their actual needs and the characteristics of the embedding table used during deployment.

\subsection{System design}\label{method_system}

Although fine-grained in-training dimension pruning has multiple benefits, it faces one 
practical challenge to realize memory saving during training: 
a straightforward implementation will
result in many small fragments of free memory. To this end, we propose
a new Virtually Hashed Physically Indexed (VHPI) embedding table design adapted from AdaEmbed~\cite{adaembed}, as illustrated in Figure~\ref{fig:system}. The original system design in AdaEmbed becomes a special case where the number of chunks is 1. 

\textbf{System Components.} The design mainly consists of a hash table, an
embedding table, and a chunk address manager. (1) VHPI hash table. Each hash table entry contains addresses of $K$ embedding chunks, $K$ utility values, and a bit mask that indicates whether the chunks have been pruned. Since $K$ is usually chosen to be small, the size of an entry is also small, which allows the hash table to have a large number of entries and a low collision rate. (2) VHPI embedding table. The embedding table stores embedding vector chunks coming from all sparse features. The size of the embedding table is set to a desired amount decided by the user, i.e., the pruning rate multiplied by the size of embeddings in an unpruned model. (3) Chunk address manager. It manages free addresses in the embedding table and maintains a free address stack which is updated with newly available chunk locations whenever a chunk is evicted during pruning. 

\textbf{Operations.} (1) Embedding access. To fetch an embedding, the sparse feature ID and the feature value are together hashed to obtain the index of an entry in the VHPI hash table. Unpruned chunks are identified by checking the $K$-bit mask, and their addresses are obtained from the entry. Embedding chunks are
then fetched from the embedding table according to the addresses. (2) Embedding eviction. When an embedding chunk needs to be evicted, its address is passed to the chunk address manager, who pushes the address into the free address stack, and the $K$-bit mask in the hash table is updated accordingly. (3) Embedding allocation. To allocate an embedding chunk, the chunk address manager fetches the next available chunk location, which is then stored in the hash table entry. Embeddings at the allocated location are initialized and the $K$-bit mask in the hash table is updated.

\begin{figure*}
%\vskip 0.2in
\begin{center}
\centerline{\includegraphics[width=0.6\linewidth]{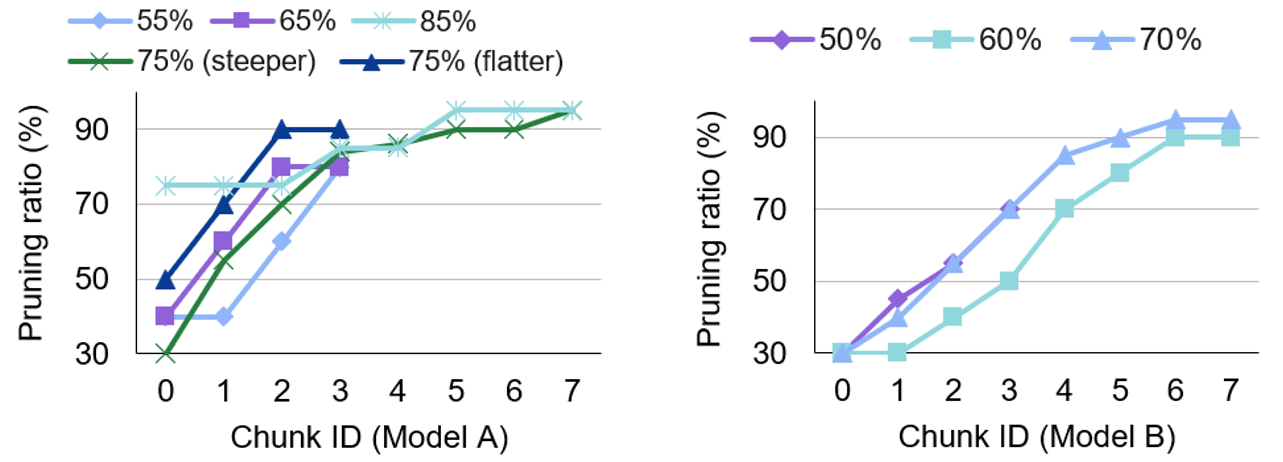}}
\caption{Per-chunk pruning ratio settings for industry models. Figure legends show the average pruning ratio for each setting. The number of chunks varies from 4 to 8. %\fan{fix the weird legend}
}
\label{fig:pruning_ratio}
\end{center}
\vskip -0.2in
\end{figure*}

\subsection{Overhead optimization}\label{method_optimization}

A major challenge in the system design described in Section~\ref{method_system} is that a naïve implementation approach will cause significant performance overhead. This overhead mainly comes from the following sources: (1) Hash table look-up. In FIITED, we need an extra mapping process to convert input indices to their actual addresses in the embedding table, or an all-zero dummy embedding if the embedding is currently pruned. (2) Multi-chunk embedding table look-up. After chunk addresses of the sparse features have been retrieved from the hash table, embedding vectors of different chunks need to be fetched from corresponding embedding tables and then aggregated to form the entire embedding vector. (3) Utility value update. After each backward process during training, the utility value of each accessed entry needs to be updated. This involves the coalescing of gradients and calculation of utility values specified in Equation ~\ref{eq:utility}. (4) Pruning overhead. In a pruning process, first, a pruning threshold has to be computed by sorting the sampled utility values and then selecting the value at a desired pruning ratio, as described in ~\ref{method_alg}. Although this overhead is already reduced by computing the threshold only on a sampled set of the whole embedding table, this process can still be expensive. Moreover, as pruning thresholds may vary between different chunks, the sampling-sorting process needs to be done for each individual chunk, further aggravating the overall time cost. After a threshold is selected, modifications need to be made on the VHPI hash tables for each chunk, evicting entries falling below the threshold and bringing back newly emerging entries.

In order to tackle the issues addressed above, we propose a set of overhead reduction methods using pipeline overlapping and parallelism techniques to optimize the performance of our system. In particular, we first use parallelism to reduce the overhead caused by the introduction of multiple chunks in our system, including embedding table look-up and pruning process. Then we further reduce the overhead by overlapping the utility value update stage of the previous batch with the forward process of the next batch. Since the calculation of utility value only requires previous frequency and gradient information, it can be overlapped with hash table look-up and embedding table look-up, reducing overhead from both sources. Furthermore, a pre-fetching technique can also be implemented in our system, where hash table look-up and embedding-table look-up can be wrapped in a dataloader to utilize the pre-fetching scheme, in which case the overhead can be further reduced by pre-fetching embeddings from in the next batch during the current training iteration.

\begin{scriptsize}
\begin{table}[t]
  \centering
  \caption{Models and datasets used for evaluation.}
  \label{table:models}
  \begin{center}
  \begin{tabular}{|l|l|l|}
    \hline
    \textbf{Model} & \textbf{Dataset} & \textbf{Embedding Table Size} \\
    \hline
    \hline
    DLRM & Criteo Kaggle & 4.02GB \\
    DLRM & Criteo Terabyte & 11.71GB\\
    DLRM & Avazu & 1.13GB \\
    DLRM & MovieLens-20M & 0.03GB \\
    \hline
    Model A (private) & private & 180GB \\
    Model B (private) & private & 430GB \\
    \hline
  \end{tabular}
\end{center}
\end{table}
\end{scriptsize}

\begin{scriptsize}
\begin{table*}[t]
  \centering
  \caption{Maximum embedding table reduction on public datasets. \\
  (ESAPN is only applicable to the MovieLens-20M dataset.)
  }
  \label{table:accuracy}
  \begin{center}
  \begin{tabular}{|l|llll|}
    \hline
    \textbf{Dataset} & \textbf{AdaEmbed} & \textbf{ESAPN} & \textbf{FIITED (Manual)} & \textbf{FIITED (Dynamic)}\\
    \hline
    \hline
    Criteo Kaggle & $50\times$ & - & $\boldsymbol{100\times}$ ($2\times$ improvement) & $\boldsymbol{100\times}$ ($2\times$ improvement) \\
    Criteo Terabyte & $66.7\times$ & - & $\boldsymbol{100\times}$ ($1.5\times$ improvement) & $\boldsymbol{100\times}$ ($1.5\times$ improvement) \\
    Avazu & $100\times$ & - & $200\times$ & $\boldsymbol{400\times}$ ($4\times$ improvement) \\
    MovieLens-20M & $1.8\times$ & $2.1\times$ & $\boldsymbol{2.7\times}$ ($1.3\times$ improvement) & $2.2\times$ \\
    \hline
  \end{tabular}
\end{center}
\end{table*}
\end{scriptsize}

\begin{figure*}
  \centering
  \begin{minipage}{0.33\linewidth}
    %\vskip 0.2in
    \begin{center}
    \centerline{\includegraphics[width=\columnwidth]{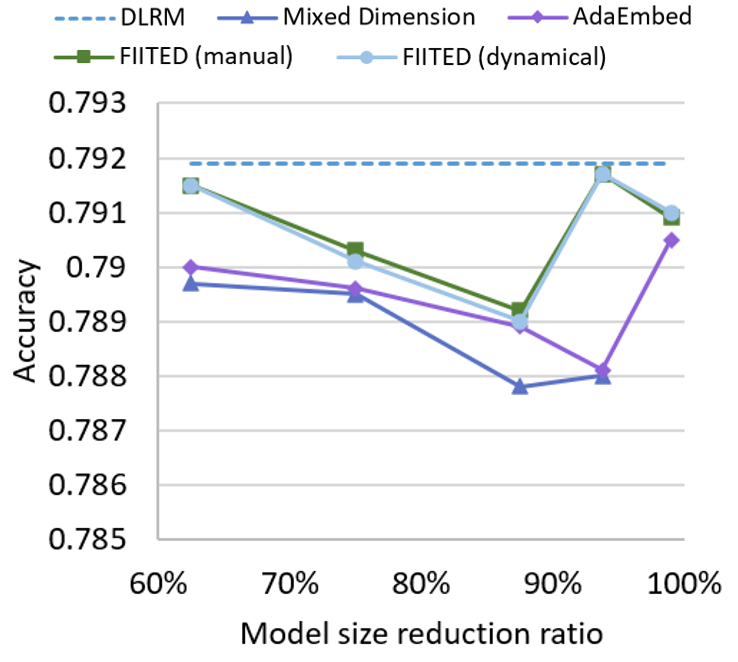}}
    \caption{Validation accuracy under different pruning ratios (62.5\%, 75\%, 87.5\%, 93.75\%, 99\%) on the Criteo Kaggle dataset. Mixed Dimension by design cannot reach 99\% pruning ratio. %\fan{fix the weird legend}
    }
    \label{fig:all_accuracy_comparison}
    \end{center}
    \vskip -0.2in
  \end{minipage}%
  \hfill
  \begin{minipage}{0.67\linewidth}
    %\vskip 0.2in
    \begin{center}
    \centerline{\includegraphics[width=\columnwidth]{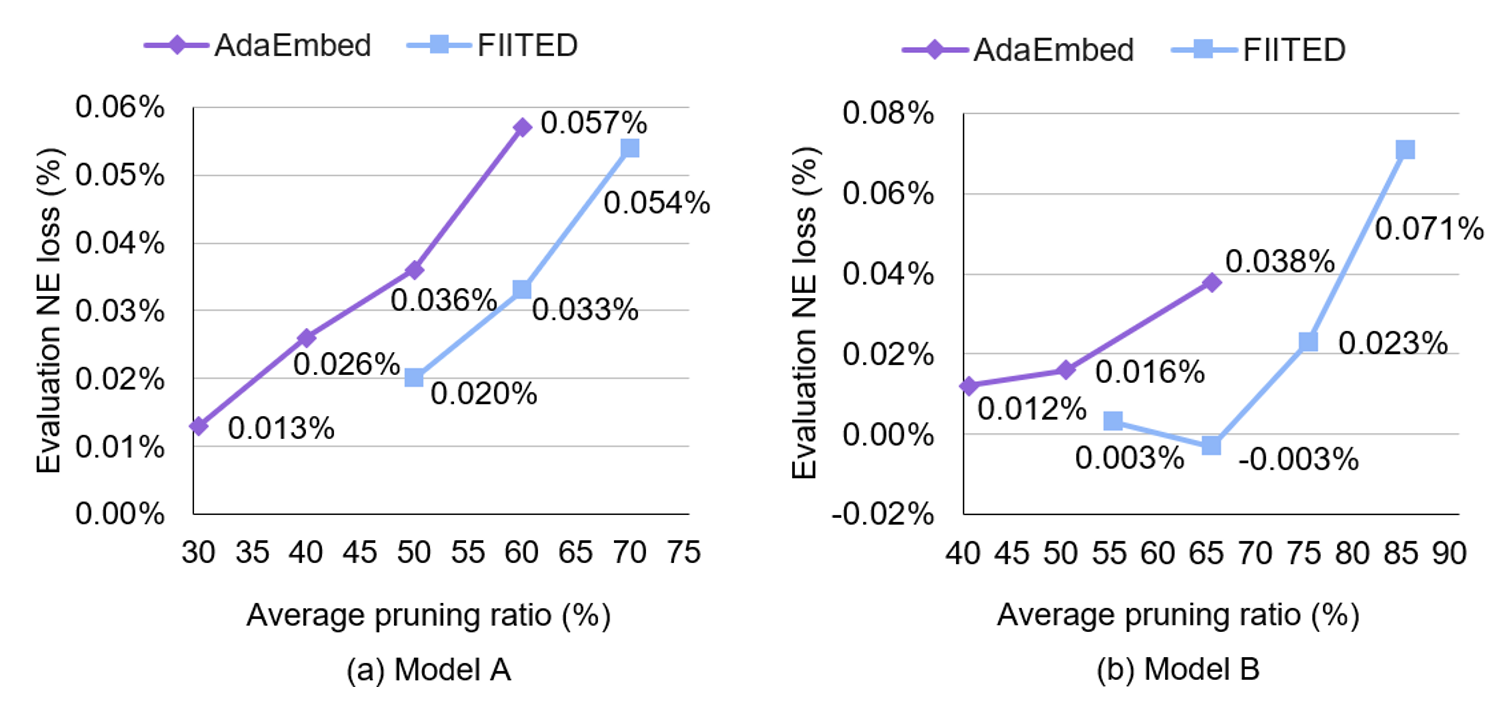}}
    \caption{Model quality vs. pruning ratio on two industry models.}
    \label{fig:rlt_AB}
    \end{center}
    \vskip -0.2in
  \end{minipage}
  \vskip -0.2in
\end{figure*}

\section{Experimental Results}\label{exp}

\subsection{Methodology}

\textbf{Baselines.} We compare FIITED with the following baselines: (1) AdaEmbed~\cite{adaembed}, a state-of-the-art \textit{in-training} embedding pruning method that prunes entire embedding rows during training, as described in Section~\ref{method_overview}; (2) ESAPN~\cite{esapn}, a state-of-the-art \textit{in-training} EDS method that selects embedding dimensions by training additional policy networks that decide to enlarge or shorten embedding dimensions during training; (3) Mixed Dimension (MD) embeddings~\cite{mix_dim}, a state-of-the-art \textit{pre-training} EDS method that selects per-feature embedding dimensions prior to training based on access frequency; (4) last but not least, the original unpruned DLRM model with uniform embedding dimensions, which will be denoted simply as DLRM in the comparison. \textbf{Baseline selection criteria:} Since the main contribution of FIITED is to save memory during training, we selected 2 \textit{in-training} embedding size optimization methods that can achieve a similar effect, in addition to 1 \textit{pre-training} method that reduces the model size before training.
We did not select \textit{post-training} EDS methods ~\cite{learn_elastic, autodim, autosrh} or other \textit{in-training} EDS methods ~\cite{twin_layer, autoemb, nis, differentiable_nis}, because they cannot reduce training memory usage, as discussed in Section~\ref{sec:related}.

\textbf{Models and datasets.} We evaluate FIITED on public models and datasets as well as private industry ones. A summary of the models and datasets can be found in Table~\ref{table:models}.
(1) Public models and datasets: We adopt an open-source DLRM framework~\footnote{\label{dlrm1} https://github.com/facebookresearch/dlrm} to train a DLRM~\cite{DLRM19} on Criteo Kaggle Display Advertising Challenge Dataset~\footnote{\label{criteo} {https://ailab.criteo.com/ressources}}, Avazu Click-Through Rate Prediction Dataset~\footnote{\label{avazu} {https://www.kaggle.com/c/avazu-ctr-prediction}} and Criteo Terabyte Dataset~\footnote{\label{terabyte}{https://labs.criteo.com/2013/12/download-terabyte-click-logs}}. We also modified this framework to train another open-source DLRM~\footnote{\label{dlrm2} https://github.com/zgahhblhc/ESAPN} used by ESAPN~\cite{esapn} on the MovieLens-20M dataset~\footnote{\label{movielens} https://grouplens.org/datasets/movielens/20m}. On the MovieLens-20M dataset, following ESAPN, we converted the multi-classification problem to a binary classification problem by viewing 4-star and 5-star reviews as positive, and others as negative. (2) Industry models and datasets: Two production models
are used in the evaluation. For ease of experimentation, both models are shrunk to 1/4 size by reducing the hash size of all sparse features by 75\%. The reduced models, named Model A and Model B, have around 180GB and 430GB sparse feature embeddings, respectively, and both contain hundreds of sparse features. They are evaluated on data generated by real-world applications. Ten consecutive days of data are used as the training set, and the following one day's data are used as the evaluation set.

\textbf{Implementation.} For evaluation using the public models and datasets, we implement FIITED as well as AdaEmbed~\cite{adaembed} on top of an open-source DLRM framework~\cite{DLRM19} based on PyTorch. Utility values are asynchronously updated with training to reduce runtime overhead, while operations for different chunks are parallelized, as described in ~\ref{method_optimization}. For evaluation on the industry models and datasets, a proof-of-concept prototype design is implemented on the company's internal deep recommendation system code base. The prototype stores both pruned and unpruned embedding chunks, and executes pruning by setting the corresponding memory regions to zero.

\textbf{Training specifications.} For evaluation using the public models and datasets, the experiments are run on 1 GPU, and we compare two methods to select per-chunk pruning ratios: manually and dynamically. To manually specify the pruning ratio for each embedding chunk, a linear function
is fitted to satisfy the desired average pruning ratio. 
We used a default sparse embedding dimension of 32 on Criteo Kaggle and Avazu datasets, 64 on Criteo Terabyte, 128 on MovieLens-20M dataset. The number of chunks $K$ is set to 2 on all datasets for FIITED.

For evaluations using the industry models and datasets, the experiments are run on 32 A100 GPUs, and the average embedding pruning ratio varies from 30\% to 85\%. 
Per chunk pruning ratios are manually chosen 
and shown in Figure~\ref{fig:pruning_ratio}. The ratios are designed 
to roughly resemble a power law distribution, which was assumed by previous EDS work ~\cite{mix_dim} and fits the observation that only a small number of ``hot" embeddings require a long dimension ~\cite{sseds}.$K$ varies between 4 and 8.
For Model A, we also experimented with 2 different pruning
ratio settings that have the same average ratio (75\%), one is “steeper” with the ratios in
different chunks span a larger range and the other is ``flatter", to investigate the effect of such a difference.
Chunk utility is simply computed as a running average of access frequency in the experiments on industry models.

\textbf{Metrics.} In this work, we mainly care about (i) maximum embedding memory saving without significantly affecting model quality. This metric showcases the core ability of FIITED by reducing the embedding table size without harming its prediction quality. (ii) overhead introduced by FIITED. As a longer end-to-end training time will cause more resource consumption, it is important to ensure that the overhead of FIITED is negligible compared with the unpruned baseline, and this overhead grows at a controllable scale as the chunk number grows.

\begin{scriptsize}
\begin{table*}[t]
  \centering
  \caption{Maximum embedding table memory saving using FIITED-2D.
  }
  \label{table:dimension_accuracy}
  \begin{center}
  \begin{tabular}{|l|lll|}
    \hline
    \textbf{Dataset} & \textbf{Entry Reduction} & \textbf{Dimension Reduction} & \textbf{Overall Reduction}\\
    \hline
    \hline
    Criteo Kaggle & $66.7\times$ & $2\times$ & $133.3\times$ \\
    Criteo Terabyte & $66.7\times$ & $2\times$ & $133.3\times$ \\
    Avazu & $100\times$ & $8\times$ & $800\times$ \\
    MovieLens-20M & $1.8\times$ & $12.5\times$ & $17.8\times$ \\
    \hline
  \end{tabular}
\end{center}
\end{table*}
\end{scriptsize}

\begin{figure*}
\begin{center}
\centerline{\includegraphics[width=0.85\linewidth]{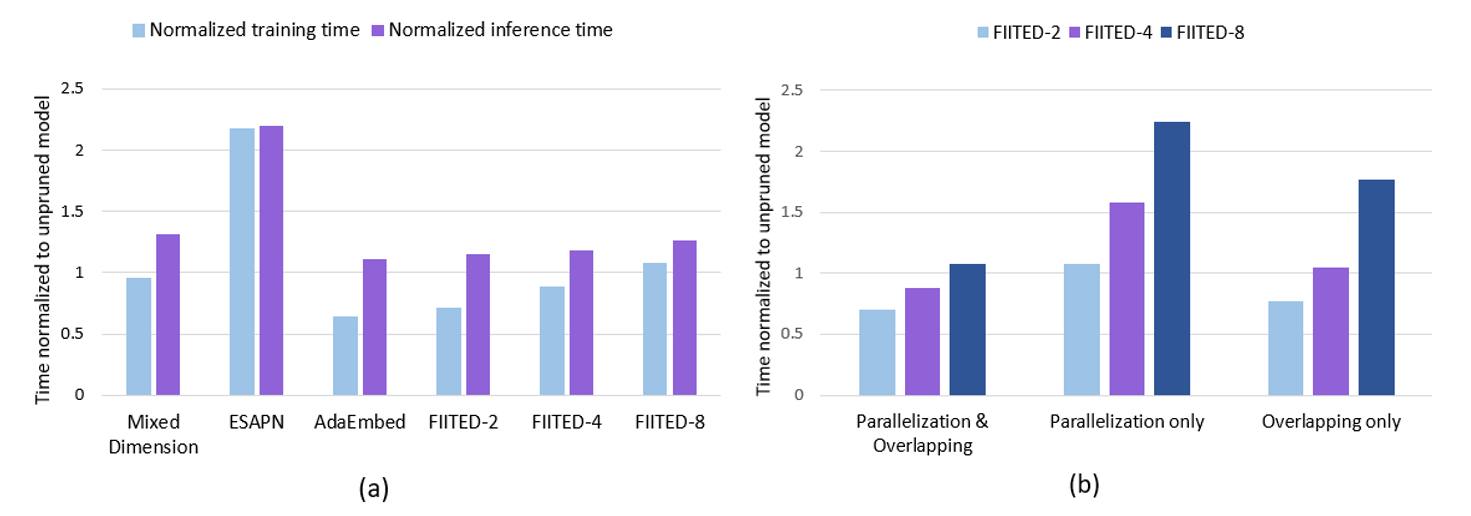}}
\caption{(a) Average training/inference time per iteration of different models. (b) Ablation study result of overhead reduction design in FIITED. Both experiments on Criteo Kaggle dataset with batch size=4096.}
\label{fig:overhead}
\end{center}
\vskip -0.2in
\end{figure*}

\subsection{Memory Saving Results}\label{exp_pruning}

In this section, we evaluate how much memory can be saved by FIITED without affecting the model quality.

\textbf{Public models.} For the public models, the criterion for model quality is prediction accuracy. According to ~\cite{zhou2018deep} and ~\cite{song2019autoint}, an accuracy loss larger than 0.1\%-level is considered significant for click-through-rate prediction tasks. Table~\ref{table:accuracy} shows the minimum equivalent number of entries that AdaEmbed, ESAPN, and FIITED are able to keep in the embedding table, while retaining the accuracy budget required by each task. On the three click-through-rate prediction datasets (i.e., Criteo Kaggle, Criteo Terabyte, and Avazu), the accuracy budget is set to less than 0.1\% lower than the unpruned model, while on the MovieLens-20M dataset, we reported the minimum number of preserved entries to obtain a comparable accuracy to ESAPN.
On all four datasets, FIITED can keep fewer embedding entries than the baselines while maintaining the model quality. Overall, FIITED is able to achieve $100\times$ to $400\times$ embedding table reduction on the three click-through-rate prediction datasets, yielding $1.5\times$ to $4\times$ improvement compared to AdaEmbed, and $2.2\times$ to $2.7\times$ on the classification dataset, a $1.05\times$ to $1.3\times$ improvement compared to ESAPN. Figure~\ref{fig:all_accuracy_comparison} shows the validation accuracy comparison of FIITED, AdaEmbed and Mixed Dimension on the Criteo Kaggle dataset under different model size reduction ratios.
FIITED consistently achieves better accuracy than the baselines, and the two pruning ratio selection approaches (manual and dynamical) yield similar results.

We then verify the performance of 2D pruning using the same public datasets mentioned above. The 2D pruning method is implemented upon the manual pruning ratio setting of FIITED. In Table~\ref{table:dimension_accuracy}, we report the minimum preserving result using the 2D pruning method of FIITED. In the table, we showcased the original number of entries and dimension size used by the four public datasets, and the average number of preserved entries and preserved dimension size used by FIITED-2D to achieve the best pruning result while meeting the accuracy requirement on each dataset, with the corresponding overall reduction ratio. Compared with the reduction results of FIITED (Manual) in Table~\ref{table:dimension_accuracy}, 2D pruning is able to achieve a higher overall reduction ratio on all four datasets, especially on Avazu and MovieLens-20M, where dimension size can be significantly reduced without harming accuracy. Overall, FIITED-2D can achieve $1.3\times$ to $6.6\times$ improvements in embedding savings compared to FIITED (Manual), and $2\times$ to $9.9\times$ improvements compared to AdaEmbed and ESAPN.

\textbf{Production models.} The criterion for model quality is Normalized Entropy (NE)~\cite{NE1,NE2} on the evaluation set; a lower NE indicates a better model. The effect of pruning is evaluated by NE loss, i.e., the percentage change in NE after model pruning compared to an unpruned model.
A positive NE loss indicates worsened model quality, and an NE loss bigger than 0.02\% is generally considered significant. 
The results are shown in Figure~\ref{fig:rlt_AB}. For both Model A and Model B, FIITED can achieve better NE than AdaEmbed given the
same average pruning ratio. For Model A, without incurring significant NE loss ($>$0.02\%), FIITED can prune 65\%-75\% of the embeddings, while AdaEmbed can only prune 50\%-65\%. For Model B, FIITED can prune up to 50\% of the embeddings, while AdaEmbed can prune
30\%-40\% of the embeddings without significant NE loss.

For Model A, we also compare two different per-chunk pruning ratio settings with the same average
ratio (75\%). The ``steeper" setting achieved a
better NE (+0.02\% VS. +0.03\%) than the ``flatter" setting. This shows that per-chunk
pruning ratios do affect model quality noticeably. Dividing a row into more chunks can
potentially enable more fine-tuned pruning ratio settings and result in better NE, but at the same
time may incur additional memory cost and access latency (Section~\ref{exp_overhead}). Setting appropriate pruning ratios and the number of chunks is a trade-off between model quality,
memory usage, and training time.

\subsection{Overhead analysis}\label{exp_overhead} 

\textbf{End-to-end training overhead.} Figure~\ref{fig:overhead}(a) shows the average training/inference time per iteration of MD embeddings, AdaEmbed, ESAPN, and FIITED with $K=2,4,8$, using Criteo Kaggle dataset with a batch size of 4096. The reported numbers are divided by the per-iteration time taken by an unpruned model for normalization purposes. As shown in the figure, during training, when $K=2,4$, FIITED is able to outperform MD embeddings and the unpruned DLRM, yielding a 1.41$\times$ and 1.13$\times$ speedup compared with the unpruned DLRM, while seeing a small overhead compared to AdaEmbed (i.e., FIITED with $K=1$) and being significantly faster than ESAPN. The performance overhead steadily grows to the number of chunks, remaining faster or negligibly slower than DLRM. During inference, the overhead of FIITED remains smaller than MD embeddings and ESAPN, and is slightly larger than AdaEmbed. Overall, the end-to-end training and inference performance of FIITED stays faster or comparable to the unpruned model (DLRM), and the overhead grows in a controllable scale to the number of chunks.

\begin{figure}[t]
\vskip 0.2in
\begin{minipage}[c]{0.48\linewidth}
\includegraphics[width=\linewidth]{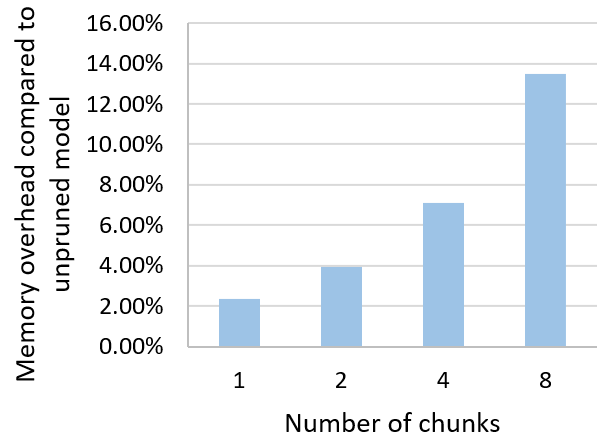}
\caption{Memory overhead of FIITED on production models.}\label{fig:sim_mem}
\end{minipage}
\hfill
\begin{minipage}[c]{0.48\linewidth}
\includegraphics[width=\linewidth]{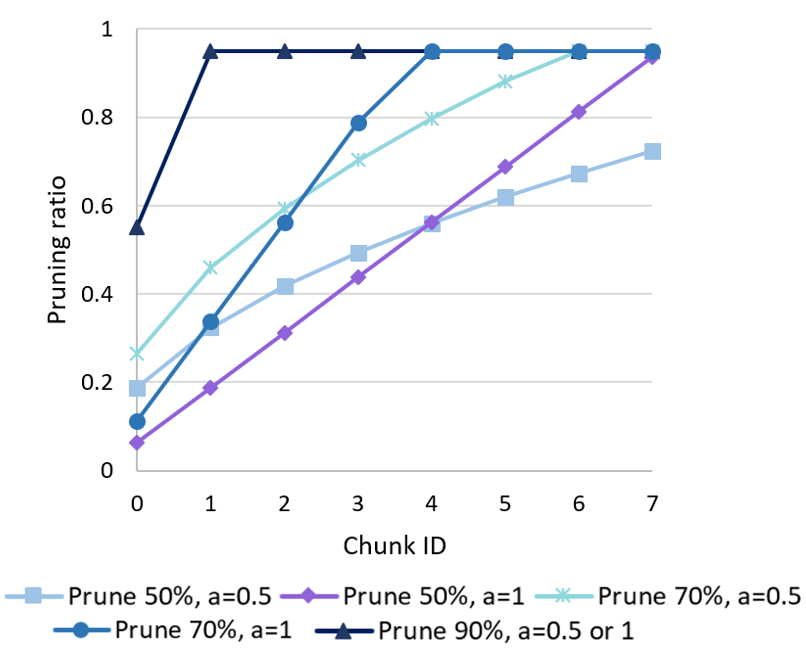}
\vskip -0.005in
\caption{Pruning ratios used in the simulation when the number of chunks is 8.
}\label{fig:sim_ratio}
\end{minipage}
\vskip -0.15in
\end{figure}

\begin{figure}[t]
\begin{center}
\centerline{\includegraphics[width=\columnwidth]{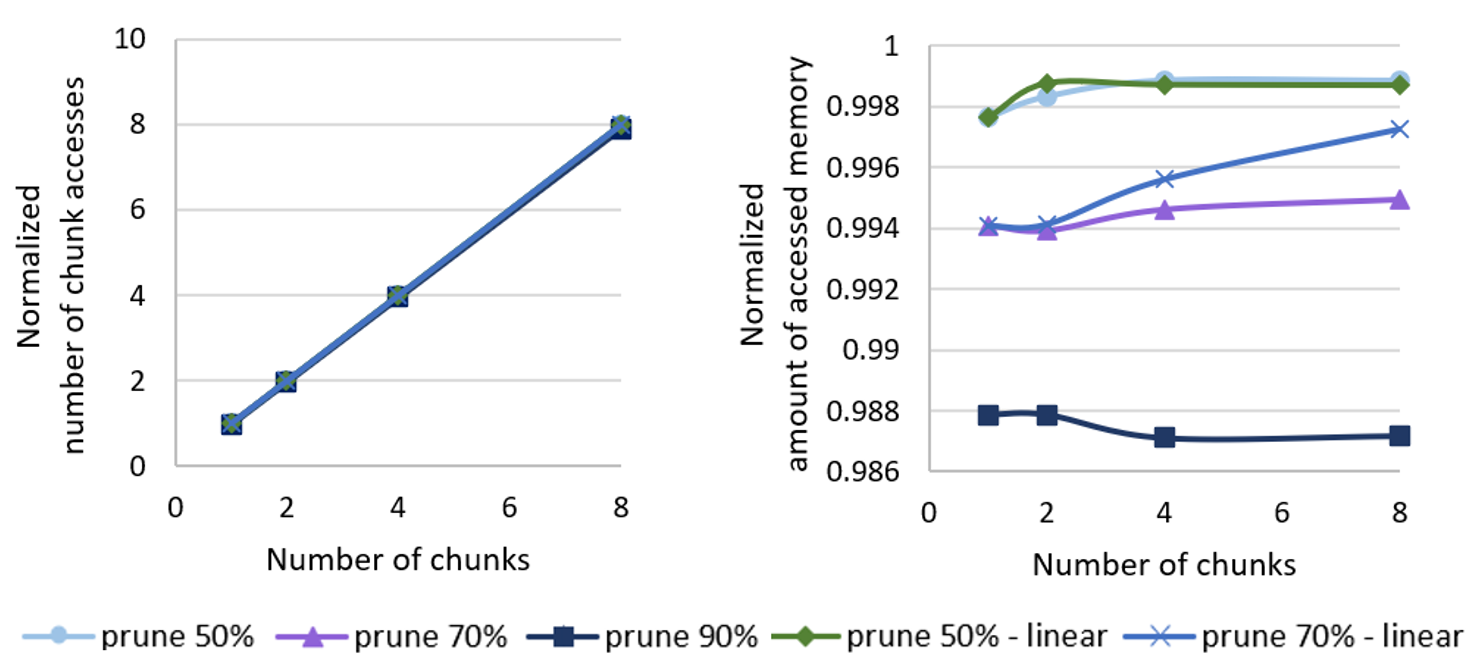}}
\caption{Embedding accesses comparison (left: number of chunk accesses; right: total amount of accessed memory, including repeated accesses). Normalization is performed by dividing values by the number of embedding accesses in an unpruned model.
}
\label{fig:sim_acc}
\end{center}
\vskip -0.2in
\end{figure}

\textbf{Memory overhead.} For a single-chunk FIITED model (i.e., $K=1$, same as AdaEmbed), the memory overhead can be estimated by $3/D$, where $D$ is the embedding dimension and 3 accounts for storing 1 chunk address, 1 utility value and 1 access frequency value per chunk. For $K>1$, the memory overhead of FIITED grows in a linear scale and becomes $3K/D$. During inference, both the utility value and the frequency value can be discarded, resulting in a $K/D$ memory overhead. To further verify the memory overhead of FIITED, we ran a simulation of the VHPI embedding store system on the Kaggle dataset, which will be discussed in detail in Section~\ref{exp_ablation}. The memory overhead of FIITED in the simulation is shown in Figure~\ref{fig:sim_mem}, where the memory overhead scales approximately linearly to the number of chunks.

\begin{figure}[t]
\begin{center}
\centerline{\includegraphics[width=\columnwidth]{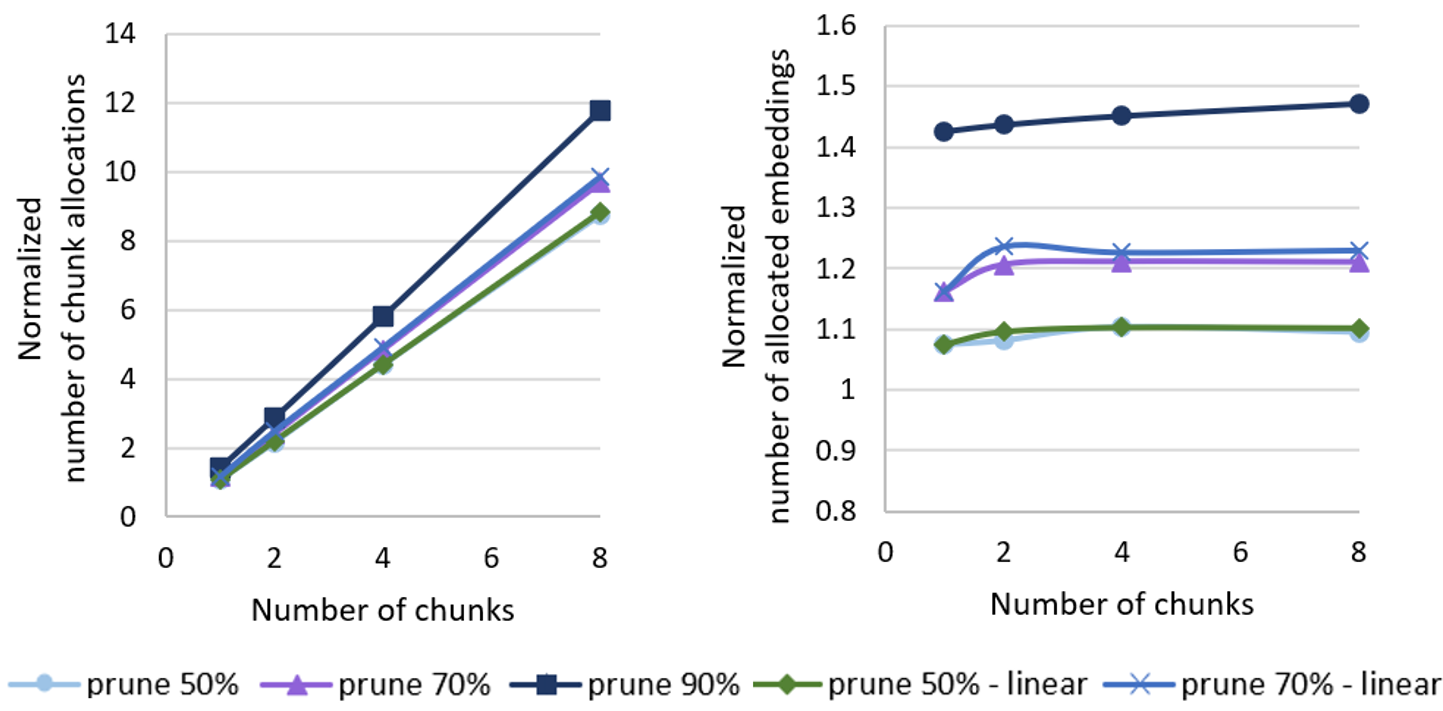}}
\caption{Embedding evictions comparison (left: number of chunk evictions; right: total amount of evicted memory). Normalization is performed by dividing values by the number of unique embeddings in an unpruned model.}
\label{fig:sim_evi}
\end{center}
\vskip -0.2in
\end{figure}

\begin{figure}[t]
\begin{center}
\centerline{\includegraphics[width=\columnwidth]{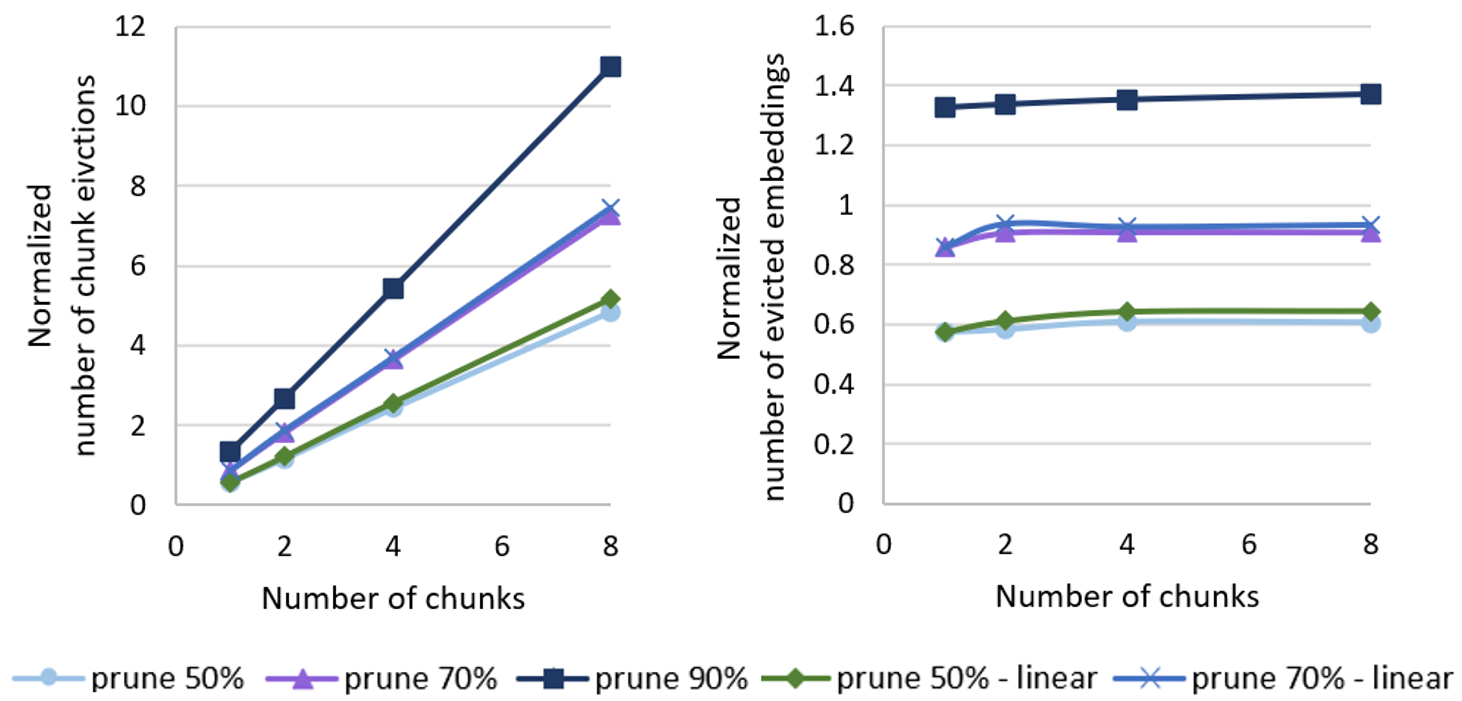}}
\caption{Embedding allocations comparison (left: number of chunk allocations; right: total amount of allocated memory, including re-allocations). Normalization is performed by dividing values by the number of unique embeddings in an unpruned model.}
\label{fig:sim_allo}
\end{center}
\vskip -0.2in
\end{figure}

\subsection{Ablation Study}\label{exp_ablation}

\begin{scriptsize}
\begin{table*}[t]
  \centering
  \caption{Comparison of existing EDS methods with FIITED.}
  \label{table:EDS_compare}
  \begin{center}
  \begin{tabular}{|l|llll|}
    \hline
    \textbf{EDS Comparison} & \textbf{Pre-training} & \textbf{Post-training} & \textbf{In-training (except FIITED)} & \textbf{FIITED} \\
    \hline
    \hline
    Require prior knowledge of training data & \textcolor{gray}{Yes} & \textbf{No} & \textbf{No} & \textbf{No} \\
    Reduce training memory usage & \textbf{Yes} & \textcolor{gray}{No} & \textcolor{gray}{No} & \textbf{Yes} \\
    Adapt to dynamic changes in data traits & \textcolor{gray}{No} & \textcolor{gray}{No} & \textbf{Yes} & \textbf{Yes} \\
    \hline
  \end{tabular}
\end{center}
\vskip -0.2in
\end{table*}
\end{scriptsize}

\textbf{Impact of training overhead optimization.}
To verify the effectiveness of our overhead optimization method proposed in Section~\ref{method_optimization}, we further conduct ablation studies on FIITED regarding the optimization effect of parallelization and pipeline overlapping. Figure~\ref{fig:overhead}(b) shows the average end-to-end per-iteration training time of FIITED with $K=2,4,8$ under the same settings as described above, compared with the system implementations with only parallelization or overlapping. All reported numbers are normalized using the time taken from the unpruned DLRM model, using the same approach as the experiment results described above. The result clearly shows the efficacy of the system design of FIITED, representing significant performance reduction compared with the system designs with only parallelization or overlapping. In the experiment of public datasets, we only used 1 GPU in order to simulate the scenario with restricted computation resources. Hence, parallelization and pipeline overlapping are only done using multi-threading and multiprocessing on a single device. On a larger distributed system, the performance overhead of FIITED can be further reduced by distributing the retrieval and utility update of different chunks on different devices, following the same scheme as shown in this section.

\textbf{Impact of VHPI embedding table.} 
In order to further analyze the overhead of FIITED, a simulation of the VHPI embedding store system is implemented in Python and run on 7 days of data in the open-source Kaggle dataset. All the VHPI embedding store operations are simulated and a full-size VHPI hash table is maintained. The VHPI embedding table is simplified in the simulation and does not contain actual embeddings. Training is not done during the simulation.

To specify the pruning ratio for each embedding chunk, a power law function $f(x) = cx^a$ at discrete points $x=(i+0.5)/K, i=0,1,2,..., K-1$ in the interval [0,1] is fitted to satisfy the desired average pruning ratio. 
We choose 2 settings for the power law parameter $a$: $a=0.5$ for a concave power function and $a=1$ for a linear power function, and compute the normalization coefficient $c$ during curve fitting. If a chunk's pruning ratio is too big (bigger than 95\%) after curve fitting, it is reduced to 95\%, and the curve fitting process is re-run for the rest of the chunks. The average pruning ratio is set to 50\%, 70\%, and 90\% in the experiments. 
The pruning ratios used in the simulation when $K=8$ are plotted in Figure~\ref{fig:sim_ratio}. 
At 90\% average pruning ratio, due to limiting the per chunk pruning ratio to a maximum of 95\%, the two power-law settings ($a=0.5$ and $a=1$) yield the same pruning ratios.

We plot the statistics of embedding accesses, evictions, and allocations in Figures~\ref{fig:sim_acc}, \ref{fig:sim_evi} and \ref{fig:sim_allo}. The left-hand sides in the figures show the total number of operations performed at the chunk level, while the right-hand sides display the averaged number of operations performed at the embedding level. The right-hand sides can be derived from the left-hand sides by dividing each value by the corresponding number of chunks. The values are normalized according to statistics of an unpruned model. 

\section{Related Work}\label{sec:related}

\subsection{Embedding Dimension Search (EDS)}\label{related_eds}
Existing EDS approaches can mainly be classified into three categories: (1) Pre-training, where embedding dimensions are decided before the actual training according to information extracted from the training dataset or a lightweight pre-training process ~\cite{sseds,pep,mix_dim}. Because the selected dimensions do not change during training, pre-training EDS misses opportunities to take advantage of changing data characteristics during training. Pre-training EDS also relies heavily on prior knowledge of the training data, which is not required by FIITED. (2) Post-training, where dimension pruning is performed after training. Additional information may be collected during training to aid pruning, and re-training is sometimes needed to boost the model performance ~\cite{learn_elastic, autodim, autosrh}. Post-training EDS cannot reduce training memory footprint. (3) In-training, where the pruning decisions are made during training. Existing works ~\cite{esapn, twin_layer, autoemb} add additional network structures on top of the original recommender model, e.g., new DNN layers, which help select embedding dimensions during training, but the majority of them still need to store embeddings at the maximum
dimension at the time of training and thus do not reduce training memory usage. One exception is ESAPN~\cite{esapn}, which stores embeddings at their current lengths during training by abandoning the old embeddings in memory whenever the embedding dimensions are increased. However, without providing 
a system design, it is unclear how ESAPN can utilize the small pieces of free memory. ESAPN also does not provide control over how much memory is used during training, while FIITED can reduce memory usage to an arbitrary desired amount. In addition, ESAPN trains one policy network per sparse feature and introduces significant training time overhead, while FIITED has negligible runtime overhead.

\subsection{Embedding Compression}

To decrease the size of embedding tables, one can: (1) reduce the size of each value in the table; (2) reduce the number of rows, i.e., the hash size; (3) reduce the number of columns, i.e., the embedding dimension; or (4) deconstruct the embedding table by replacing the traditional 2D array storage format with novel designs. (1) is commonly achieved by quantization ~\cite{quant1,quant2,quant3}. (2) involves designing novel hashing methods to reduce the number of rows while maintaining the model quality ~\cite{hash1,hash2,hash3}. (3) has been discussed previously in Section~\ref{related_eds} and includes EDS methods performed prior to, during or after training. 
As shown in Table~\ref{table:EDS_compare}, unlike existing EDS methods, FIITED is able to reduce the training memory footprint while adjusting embedding dimensions based on dynamic changes in data characteristics.
For (4), novel embedding storage designs in existing research include constructing embeddings from two separate tables ~\cite{onthefly}, via multi-layer embeddings ~\cite{multilayer}, and by applying transformation matrices to a small set of anchor embeddings ~\cite{anchor}.

\section{Conclusion}

In this paper, we propose FIITED, an in-training embedding dimension optimization method that is able to directly cut down the training memory footprint of DLRMs. Given a memory budget, FIITED can be plugged directly into training without any need for prior knowledge of training data, pre-training, or re-training. Embedding dimensions are adjusted during training in a fine-grained manner while changing data statistics are taken into consideration. Experiments on two industry models show that FIITED consistently achieves better NE than
a state-of-the-art in-training embedding pruning method given the same average pruning ratio, and can prune
much more than the baseline (65\% vs. 50\% for Model A, 50\% vs. 30\% for
Model B) without affecting evaluation NE. On public datasets, FIITED can reduce the embedding size by 100$\times$ to 800$\times$ on three click-through-rate datasets and $2.2\times$ to $17.8\times$ on one classification dataset, without significant accuracy loss.

%%%%%%% -- PAPER CONTENT ENDS -- %%%%%%%%

%%%%%%%%% -- BIB STYLE AND FILE -- %%%%%%%%
\bibliographystyle{IEEEtranS}
\bibliography{refs}
%%%%%%%%%%%%%%%%%%%%%%%%%%%%%%%%%%%%

\end{document}